\RequirePackage{fix-cm}
\documentclass[smallextended]{svjour3}       
\smartqed  
\usepackage{graphicx}
\usepackage{caption}
\usepackage{nicefrac}
\usepackage{amsfonts}
\usepackage[T1]{fontenc}
\usepackage{xspace}
\usepackage{flushend}
\usepackage{ifthen}
\usepackage{color}
\usepackage{url}
\usepackage{multirow}
\usepackage{listliketab}
\usepackage{amssymb}
\usepackage[linesnumbered,lined,boxed,commentsnumbered]{algorithm2e}
\usepackage{algorithmic}
\usepackage{booktabs}
\usepackage{comment}
\usepackage{amsmath}
\usepackage{bbding}
\usepackage{listings}
\usepackage{float}
\usepackage[misc]{ifsym}
\usepackage{tabularx}
\usepackage{cite}
\usepackage{rotating}
\usepackage{xcolor}
\usepackage{natbib}
\usepackage{tcolorbox}
\usepackage{ulem}
\usepackage{censor}
\usepackage{subcaption}
\usepackage{pgfplots}
\pgfplotsset{compat=1.14}
\usepackage{filecontents}
\usepackage{pgfplotstable}

\def\mybar#1{
  #1s & {\color{grey}\rule{#1cm}{8pt}}}
\definecolor{Fork}{HTML}{99CCFF}
\definecolor{Non-Fork}{HTML}{FFCCCB}

\newcommand{\mycolor}[1]{{\color{black}{#1}}}
\newcommand{\newcomer}{\mycolor{Newcomer OSS-Candidate}}
\newcommand{\newcomers}{\mycolor{Newcomer OSS-Candidates}}

\newcommand\wang[1]{{\textcolor{purple}{#1}}}
\def\mybar#1{
  {\color{gray}\rule{#1cm}{8pt}}}

\newcommand{\RqOne}{($RQ1$)}
\newcommand{\RqTwo}{($RQ2$)}
\newcommand{\RqThree}{($RQ3$)}
\newcommand{\RqFour}{($RQ4$)}

\newcommand{\RqOneSen}{{\RqOne~What kinds of repositories does a Newcomer OSS-Candidate target?}}

\newcommand{\RqTwoSen}{\RqTwo~What are the kinds of first contributions that come from \newcomers?}

\newcommand{\RqThreeSen}{\RqThree~To what extent do \newcomers~practice social coding with their first contributions?}

\newcommand{\RqFourSen}{\RqFour~What is the proportion of \newcomers~that eventually onboard an OSS project?}

\colorlet{punct}{red!60!black}
\definecolor{background}{HTML}{EEEEEE}
\definecolor{delim}{RGB}{20,105,176}
\colorlet{numb}{magenta!60!black}

\lstdefinelanguage{json}{
    basicstyle=\small,
    numbers=left,
    numberstyle=\scriptsize,
    stepnumber=1,
    numbersep=8pt,
    showstringspaces=false,
    breaklines=true,
    frame=lines,
    backgroundcolor=\color{background},
    literate=
     *{0}{{{\color{numb}0}}}{1}
      {1}{{{\color{numb}1}}}{1}
      {2}{{{\color{numb}2}}}{1}
      {3}{{{\color{numb}3}}}{1}
      {4}{{{\color{numb}4}}}{1}
      {5}{{{\color{numb}5}}}{1}
      {6}{{{\color{numb}6}}}{1}
      {7}{{{\color{numb}7}}}{1}
      {8}{{{\color{numb}8}}}{1}
      {9}{{{\color{numb}9}}}{1}
      {:}{{{\color{punct}{:}}}}{1}
      {,}{{{\color{punct}{,}}}}{1}
      {\{}{{{\color{delim}{\{}}}}{1}
      {\}}{{{\color{delim}{\}}}}}{1}
      {[}{{{\color{delim}{[}}}}{1}
      {]}{{{\color{delim}{]}}}}{1},
}

\begin{document}

\title{Newcomer OSS-Candidates: Characterizing Contributions of Novice Developers to GitHub}

\author{IFraz Rehman \Letter\and
        Dong Wang\and
        Raula Gaikovina Kula\and
        Takashi Ishio\and
        Kenichi Matsumoto
}

\institute{
    \Letter~Corresponding author - IFraz Rehman
    \at Nara Institute of Science and Technology, Japan\\
    \email{\{rehman.ifraz.qy4\}@is.naist.jp}
     \and
    Dong Wang
    \at Kyushu University, Japan\\
    \email{wang.dong.vt8@is.naist.jp}\\
    Raula Gaikovina Kula, Takashi Ishio, Keniichi Matsumoto 
    \at Nara Institute of Science and Technology, Japan\\
    \email{\{raula-k,ishio,matumoto\}@is.naist.jp}
}

\date{Received: date / Accepted: date}

\maketitle

\begin{abstract}
The ability of an Open Source Software (OSS) project to attract, onboard, and retain any newcomer is vital to its livelihood. 
Although, evidence suggests an upsurge in novice developers joining social coding platforms (such as GitHub), the extent to which their activities result in a OSS contribution is unknown.
Henceforth, we execute the protocols of a registered report to study activities of a  ``\texttt{\newcomer}'', who is a novice developer that is new to that social coding platform, and has the intention to later onboard an OSS project.
Using GitHub as a case platform, we analyze 171 identified \newcomers~to characterize their contribution activities.
Results show that \newcomers~are likely to target software based repositories (i.e., 66\%), and their first contributions are mainly associated with development (commits) and maintenance (PRs).
\newcomers~are less likely to practice social coding, but eventually end up onboarding (i.e., 30\% quantitative, 70\% follow-up survey) an OSS project.
Furthermore, they cite \textit{finding a way to start} as the most challenging barrier to contribute.
Our work reveals insights on how newcomers to social coding platforms are potential sources of OSS contributions.

\keywords{Newcomers, Open Source Software, GitHub}
\end{abstract}

\section{Introduction}
\label{sec:introduction}

The success of Open Source Software (OSS) has always been based on the continuous influx of newcomers and their active involvement~\citep{park2009VISSOFT}.
Previous studies have shown evidence that many contemporary projects are at risk of failure, with one of the reasons being the inability to attract and retain newcomers~\citep{Fang2009JMIS,Valiev2018FSE}. 
For example,~\cite{Coelho2017FSE} proposed two strategies that include newcomers which aim to transfer the project to new maintainers and to accept new core developers.
In another study,~\cite{Steinmacher2014} presented a model that analyzes the influential forces to newcomers being drawn or pushed away from a project.
In contrast, the rise of social coding platforms has led to an explosion of potential developers.
GitHub reported\footnote{Statistics from \url{https://octoverse.github.com} accessed January 2020} around 10 million-plus new users in 2020 and allows over 40 million developers to showcase their skills to the world's largest community (44 million upstream repositories).
With this upsurge in user activity, 
However, the extent to which these developers activities result in a contribution to OSS projects is unknown.

The term {newcomer} has usually been used in a loose way in literature~\citep{Steinmacher2014}.
Inspired by the incubation of OSS projects on GitHub, we coin the term ``Newcomer OSS-Candidate'', \textit{who is not yet a newcomer, but has potential to become one}. Concretely, we define a Newcomer OSS-Candidate as a developer that satisfies these three criteria: 1) is a developer that does not have any prior experience contributing to an OSS project, 2) is a new user to a social coding platform, and 3) has the intention to onboard an OSS project hosted on a social coding platform.
Although there is a complete body of work that has studied the barriers and struggles of newcomers~\citep{Steinmacher2014, Steinmacher2015ICSS}, none has explored the contribution kinds of \newcomers. Most of the work revolves around newcomers that have already onboarded OSS projects.

This study is an execution of the protocol reported by~\citet{ifrazNC2020}, using GitHub as a case platform.  
We studied 171 \newcomers~and their GitHub repositories, guided by four research questions: 

\begin{itemize}
    \item \textbf{\RqOneSen}
    \cite{kalliamvakou2014MSR} showed that most repositories hosted on GitHub are non-software. 
    However, since Newcomer OSS-Candidates have the intention to later onboard a software project, we would like to test the assumption that \textit{(H1) Newcomer OSS-Candidates are more likely to target software repositories}. Since GitHub users can either create their own upstream repositories or fork existing repositories, we compare these two kinds of repositories. \\
    We observe that 66\% of Newcomer OSS-Candidates target software based repositories.
    The statistical test indicates that hypothesis \textit{H1} is established.
    Furthermore, Experimental and Documentation are the most frequently targeted software repository kinds for fork and upstream repositories, i.e., 24\% and 21\%, respectively.
    \item\ \textbf{\RqTwoSen}
    \cite{Hattori2008ASE} showed that OSS projects constantly add new content to software (i.e., development) more frequently than maintaining existing code. Hence, for this RQ, our motivation is to understand whether or not Newcomer OSS-Candidates are more likely to add new content or maintain the repository. 
    Hence, by studying these two types of contributions, we test the hypothesis that \textit{(H2) Contributions to GitHub repositories from Newcomer OSS-Candidates are more likely to do development activities.} 
    We analyze two kinds of GitHub contributions, either a direct contribution through a commit, or a submitted Pull Request (PR).\\
    For the first commit contributions, we find that 74\% of contributions from Newcomer OSS-Candidates are related to development activities.
    For the first PR contributions, our results show that 60\% of contributions are associated with management activities. 
    The statistical tests confirm that our hypothesis \textit{H2} is established in first commit contributions, while is not established in first PR contributions.
    \item \textbf{\RqThreeSen}
    Since GitHub is a social coding platform, we would like to explore the extent to which a Newcomer OSS-Candidate is likely to make a social contribution as their first contribution.
    Specifically, we analyze whether or not a Newcomer OSS-Candidate shares code, which is measured by single or multiple authorship on a file.
    Hence, similar to RQ3, we explore the commit and PR contributions to test the hypothesis \textit{(H3) Newcomer OSS-Candidates are more likely to contribute to a file with multiple authorship.} \\
    Our results show that after joining GitHub, a majority of \newcomers~(i.e., 73\% of first commits and 59\% of PRs) do not share code with other authors.
    Moreover, the statistical tests validate that our hypothesis \textit{H3} is not established for both first commit and first PR contributions.
    \item \textbf{\RqFourSen}
    In accordance with our definition, we explore the extent to which these Newcomer OSS-Candidates eventually onboard an OSS project.
    We would like to explore the proportion of Newcomer OSS-Candidates who eventually onboard an OSS project.
    Additionally, we validate what kinds of barriers that Newcomer OSS-Candidates face when onboarding OSS repositories.\\
    Our quantitative analysis shows that 30\% of Newcomer OSS-Candidates eventually onboarded engineered OSS repositories. Complementary, a follow-up user survey shows that 70\% of studied participants ended up making contributions to an OSS repository. Newcomer OSS-Candidates strongly agreed that they face the barrier of \textit{finding a way to start}, while \textit{social interaction} received the most mixed responses as a barrier.
\end{itemize}


The remainder of this paper is organized as follows: Section 2 describes the identification procedure for \newcomers.
Section 3 reports the approaches and results of our empirical study, while Section 4 discusses the deviations, lesson learned and our findings. 
Section 5 discloses the threats to validity, Section 6 presents related work and finally, we conclude the paper in Section 7. 
To facilitate replication and future work in the area, we have prepared a replication package, which includes the studied 171 Newcomer OSS-Candidates' repositories, manually labeled datasets, the scripts for the quantitative analyses, and the survey materials.
The package is available online at \url{https://github.com/NAIST-SE/NewcomerCandidate}.

\section{Identifying Newcomer OSS-Candidates}
\label{sec:preliminary_survey}

\begin{table*}[t]
\centering
  \caption{Survey Questions sent to potential respondents}
  \label{tab:data_col1}
  \begin{tabular}{lp{4.5cm}}
    \toprule
    Survey Questions for \newcomer\\ 
    \midrule
    Q1) What is your motivation to make a contribution to GitHub?\\
    (a) Learning to Code. \\ 
    (b) Assignment or Experiment Project.\\
    (c) Intend to contribute to an Open Source. \\
    (d) Use to showcase my programming skills.\\
    (e) Others.\\\hline
    Q2) Did you have prior experience contributing to an OSS before GitHub?\\
    (Yes/No)\\
    \bottomrule    
\end{tabular}
\end{table*}
In this section, we describe the process of identifying Newcomer OSS-Candidates.
As per our registered report \citep{ifrazNC2020}, we used the first-contribution community\footnote{\url{https://github.com/firstcontributions/first-contributions/blob/master/Contributors.md}} in GitHub as our data source for collecting Newcomer OSS-Candidates. 
The community is an initiative established to help beginners make their first contributions on GitHub and currently has over 5,000 contributors, over 39.7 thousand forks, and over 21 thousand stars as of October 2021. 
To extract the survey respondent candidates, we used command \texttt{"git log --pretty=format:\%ae"}\footnote{\url{https://git-scm.com/docs/pretty-formats}} on Contributors.md file provided by the community and were able to get 17,507 respondent candidates.
We sent our online survey invitation\footnote{\url{https://tinyurl.com/r7acxvn}} to reach up to 4,000 respondent candidates through email and a slack channel.\footnote{\url{https://firstcontributions.slack.com/}}
Our survey was open from March 3, 2020 to March 31, 2020 (around a four-week period).
We received 208 responses, allowing us to mine their repositories and contributions by providing their GitHub IDs.
In the survey, we validate the definition of our Newcomer OSS-Candidate by asking two questions. 
The two questions are presented in Table~\ref{tab:data_col1}. 
Besides, respondents were also asked about their interests, and their perception rank of their programming skills. 

\begin{table}[t]
\caption{Two questions in our survey}
\fontsize{9}{11}\selectfont
\tabcolsep=0.5cm
\begin{subtable}[h]{\textwidth}
\centering
\begin{tabular}{lrl}
\toprule
Have you had any prior OSS experience? & Percent &  \\ \midrule
No    &   85\% & \mybar{0.85}   \\
Yes   &    15\% &\mybar{0.15}   \\
\bottomrule
\end{tabular}
        \caption{Answers to Q1 of the survey}
        \label{tab:prior_exp}
     \end{subtable}
\begin{subtable}[h]{\textwidth}
\centering
\begin{tabular}{lrl}
\toprule
What is the motivation to contribute? & Percent &  \\ \midrule
(a) Learning to Code.    &   58\% & \mybar{0.58}   \\
(b) Assignment or Experiment Project.   &    21\% &\mybar{0.21}   \\
(c) Intend to contribute to an Open Source. &    82\% & \mybar{0.82} \\
(d) Use to showcase my programming skills. &    42\% & \mybar{0.42} \\
(e) Others &    5\% & \mybar{0.05} \\
\bottomrule
\end{tabular}
       \caption{Answers to Q2 of the survey}
       \label{tab:motivation}
    \end{subtable}
\\
     \label{tab:surveyData}
\end{table}

\textbf{171 Identified \newcomers.} Table~\ref{tab:surveyData} presents the survey answers that are related to prior OSS experience of respondents and their motivations to contribute. 
Table~\ref{tab:motivation} shows that 82\% of respondents (i.e., 171 responses) intend to contribute to an OSS project. 
Furthermore, these respondents claim that they have not had any prior OSS experience.
Henceforth, according to our definition of Newcomer OSS-Candidate that is described in the Introduction, we used these 171 participants to further track their repositories and contributions for our subsequent analyses.\\

\section{Findings}
\label{sec:Study_design}
We follow the protocol that is highlighted in our registered report \citep{ifrazNC2020} to answer all RQs. Each research question comprises of the approach and their results. 
Deviations to the protocol are highlighted in Section 4.1 (Discussion).
\label{sec:repo_target}
\subsection{Target Repositories (RQ1)}

\textit{Approach.} To answer RQ1, we first construct the \textit{(D1) \newcomer~Repository Dataset}, which is a mapping of our selected \newcomer~information (as described in Section 2) with their GitHub repository contributions.
Using the GitHub REST API (GitHub, 2020) and the credentials of the 171 survey participants, we retrieved 2,392 unique contributed repositories, consisting of 936 fork\footnote{\url{https://docs.github.com/en/get-started/quickstart/fork-a-repo}} and 1,456 upstream\footnote{\url{https://docs.github.com/en/get-started/quickstart/github-glossary#upstream}} repositories.
Under the guidance of \cite{Borges2016ICSME, kalliamvakou2014MSR}, we classify the repositories into software and non-software. 
The definitions of software and non-software repositories are described below:
\begin{itemize}
    \item{\textit{(Software) Application Software:}} systems that provide functionalities to end-users, like browsers and text editors. 
    \item{\textit{(Software) System Software:}} systems that provide services and infrastructure to other systems, like operating systems, middleware, servers, and databases. 
    \item{\textit{(Software) Web libraries and frameworks.}}  
    \item{\textit{(Software) Non-web libraries and frameworks.}} 
    \item{\textit{(Software) Software tools:}} systems that support software development tasks, like IDEs, package managers, and compilers. 
    \item{\textit{(Software) Documentation:}} repositories with documentation, tutorials, source code examples. 
    \item{\textit{(Software) Experimental:} repositories include demos, samples, test code, and tutorial examples.}
    \item{\textit{(Non-Software) Storage:}} category includes repositories documents and files for personal use, such as presentation slides, resumes, e-books, music files etc.
    \item{\textit{(Non-Software) Academic:}} class and university research projects come under this category. \item{\textit{(Non-Software) Web:}} under this category we classify websites and blogs.
    \item{\textit{(Others) No longer accessible/Empty:}} repositories that gave 404 error, containing only a license file, a gitignore file, a README file, or no files at all were placed under this category.
\end{itemize}

As per the registered report, we use a qualitative method to manually classify the different kinds of repositories.
Following the protocol, with a confidence level of 95\% and a confidence interval of 5\footnote{https://www.surveysystem.com/sscalc.htm}, we draw a statistically representative sample from \textit{(D1)} to end up with 273 fork repositories and 304 upstream repositories.
To evaluate the validity of our manual coding, we randomly selected 30 repositories from the representative sample, and then the first three authors independently coded these repositories. 
The three authors then measured the inter-rater agreement using Cohen’s Kappa~\citep{viera2005understanding} as the measure of agreement.
In the end, the Kappa agreement for fork repositories was nearly perfect (i.e., 0.91), while the score for upstream repositories was substantial (i.e., 0.76).
Based on this encouraging result, the first author then completed the manual coding for the rest of the representative sample.

For our significance testing, different from the registered report\footnote{ Please refer to Section 4.1 for deviations explanations}, we validate our hypothesis \textit{(H1) \newcomers~are more likely to target software repositories},  using the one proportion Z-test~\citep{z_test} as it compares an observed proportion to a theoretical one when the categories are binary.
\begin{table}[t]
    \centering
    \caption{Proportion of software and non-software repositories targeted by \newcomers. Around 66\% of \newcomers~target Software repositories.}
    \label{tab:softnonsoft}
    \begin{tabular}{|c|c|c|} 
    \hline
    Category & Percent (\%) & Fork \& Upstream (\%) \\ 
    \hline
    \multirow{2}{*}{Software} & \multirow{2}{*}{66} & Upstream (52)~ \\ 
    \cline{3-3}
     &  & Fork (48) \\ 
    \hline
    \multirow{2}{*}{Non-Software} & \multirow{2}{*}{24} & Upstream (55) \\ 
    \cline{3-3}
     &  & Fork (45) \\
    \hline
    \multirow{2}{*}{Others} & \multirow{2}{*}{10} & - \\ 
    \cline{3-3}
     &  & - \\
     \hline
    \end{tabular}
\end{table}

\textbf{Proportion of Software and Non-Software Repositories.}
Table~\ref{tab:softnonsoft} shows the proportion of software and non-software
based repositories that \newcomers~target.
We see that 66\% of \newcomers~target repositories are software based and follow sound software engineering practices in each dimensions.
Furthermore, \newcomers~are less likely to target non-software based repositories, accounting for 24\%. Specifically, we observe that 10\% of repositories are classified as Others.
Through the manual analysis, these repositories are either ``No longer accessible'' or ``Empty''.
Upon in-depth analysis of repositories (i.e., Fork and Upstream), we observe that the dominant repositories for software and non-software are upstream i.e., 52\% and 55\%.

\begin{figure}
\centering
\includegraphics[width=\linewidth]{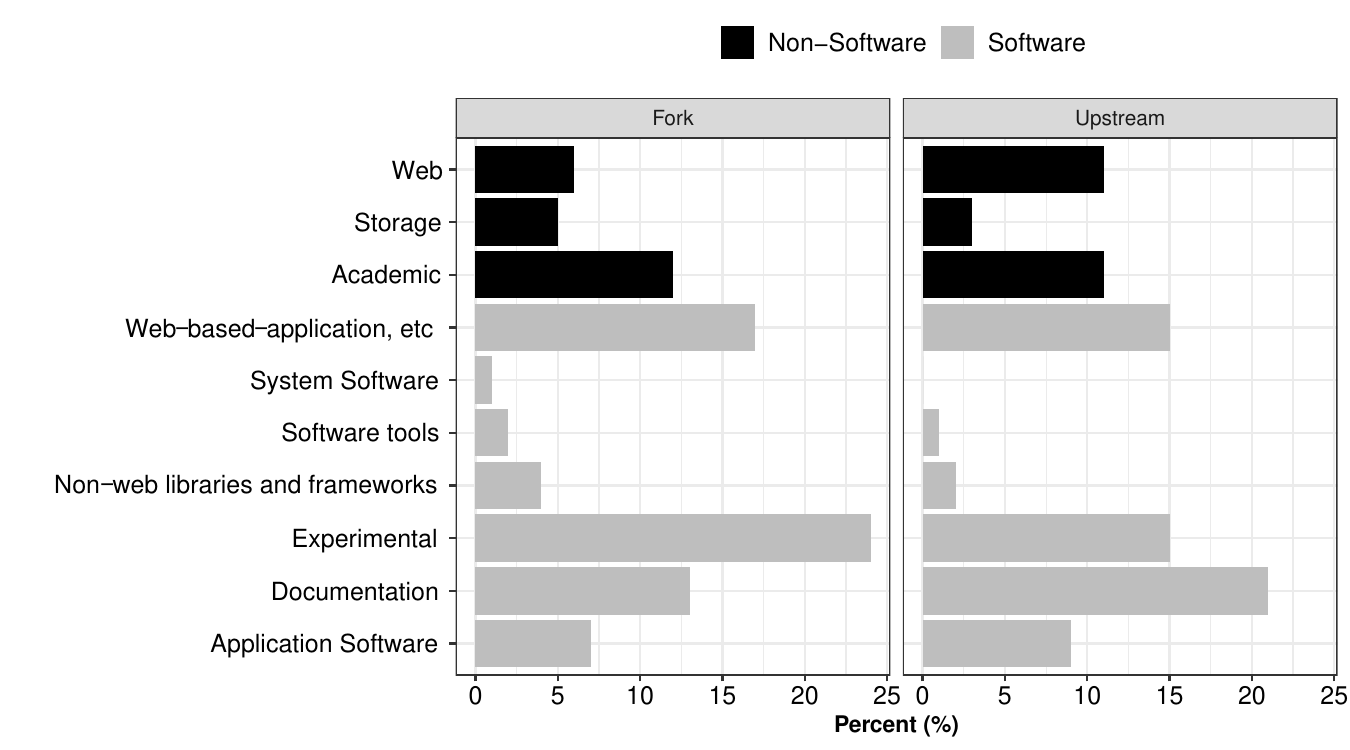}
\caption{Frequency for contributed repository kinds with Fork and Upstream. Experimental and Documentation are the most frequently targeted software repository kinds, i.e., 24\% and 21\%, respectively.}
\label{fig:rq3}
\end{figure}

\textbf{Frequency of Contributed Repository Kinds.} Figure~\ref{fig:rq3} shows that Documentation (21\%), Experimental (15\%), Web-based-applications, libraries and frameworks (15\%) are the most frequently targeted upstream software repositories kinds.
The other kinds of repositories that \newcomers~frequently target are Academic (12\%), Web (11\%), and Application Software (9\%).
On the other side, we find that Experimental (24\%) and Web-based-application, libraries, and frameworks (17\%) are the most commonly targeted fork repositories kinds.
The other kinds of fork repositories commonly targeted are Documentation (13\%) and Academic (12\%).

Our statistical test validates a significant difference between the proportion of software and non-software repositories that \newcomers~target, with a p-value $<$ 0.001.
The result indicates that our proposed hypothesis, i.e., \textit{(H1) \newcomers~are more likely to target software repositories}, is established.

\begin{tcolorbox}
    \textbf{RQ1 Summary:} Results show that 66\% of \newcomers~target software based repositories. Our proposed hypothesis that  \textit{(H1) \newcomers~are more likely to target software repositories} is established. Furthermore, Experimental and Documentation are the most frequently targeted software repository kinds for both fork and upstream repositories with 24\% and 21\%, respectively.
\end{tcolorbox}
\label{sec:kinds_of_contribition}
\subsection{Kinds of Contributions (RQ2)} 

\textit{Approach.}
To answer RQ2, different from the registered report, we analyze the first contributions with two types, i.e., first commit and first PR. As such, we constructed a new dataset from RQ1, which is \textit{(D2) First Contribution Dataset}.
To do so, we first obtain the earliest GitHub repositories each of the 171 \newcomers.
For the quality purpose, we ignore any \textit{test and not meaningful} commits by filtering out experimental repositories that have been identified in RQ1.
Furthermore, from our initial list of 171 participants, we remove another five participants. Three participants had not made any contributions to their fork or upstream repositories, and another two participants had become inactive since the initial survey.
Hence, we ended up with a total of 166 first commits and 97 PRs from 166 \newcomers.
As per the registered report, we then classify the contributions according to \cite{Hattori2008ASE}:
\begin{itemize}
    \item{\textit{Development (forward engineering and non-software):}} based on the forward-engineering type proposed by Hattori and Lanza (2008), the development activities relate to incorporation of new features and implementation of new requirements for both software and non-software. Examples of development for non-software repositories include adding new content for websites or documentation. 
    \item{\textit{Repository Initializing (sub-category of development):}} derived from the forward-engineering category, we identify any first commits as the initializing commits to a new repository.
    \item{\textit{Re-engineering:}} maintenance activities are related to refactoring, redesign and other actions to enhance the quality of the code without properly adding new features.
    \item{\textit{Corrective Engineering:}} maintenance activities
    handle defects, errors and bugs in the software.
    \item{\textit{Management:}} maintenance activities are those unrelated to codification, such as formatting code, cleaning up, and updating documentation.
\end{itemize}

To validate the understanding of the taxonomy of contribution kinds,
we randomly selected 30 contributions of first commits and PRs, and then the first three authors independently coded these contributions, similar to RQ1. Since \cite{Hattori2008ASE} used a set of keywords, we applied the keywords as an initial guide. However, when deciding the classification, we consider the commit and PR attributes (i.e., title, message, and description) to have a better understanding of the context.
Similar to RQ1, we use Cohen’s Kappa. The Kappa agreement scores for classifying contribution kinds of first commits and PRs were both substantial (i.e., 0.72 and 0.79, respectively).  After the agreement measurement, the first author then completed the remaining sample.

To validate our hypothesis \textit{(H2) Contributions to GitHub repositories from \newcomers~are more likely to do development activities}, similar to RQ1, we use the one proportion Z-test~\citep{z_test}. 
To fit the formula of the statistical test, we merge \textit{Development} and \textit{Repository Initializing} into the \textit{Development} category, and we merge \textit{Re-engineering}, \textit{Corrective Engineering}, and \textit{Management} into the \textit{Maintenance} category.

\begin{table}[t]
\centering
\caption{Frequency for Contribution's Kinds of \newcomers. In the first commits, 43\% of \newcomers~are typically engaged in repository initializing activities, and 60\% are engaged in the management activities of the PRs.}
\label{fig:rq2_graph}
\begin{tabular}{llccc} 
\toprule
\begin{tabular}[c]{@{}l@{}}\textbf{First}\\\textbf{Contributions}\end{tabular} & \textbf{Kinds} & \begin{tabular}[c]{@{}c@{}}\textbf{Percent }\\\textbf{(\%)}\end{tabular} & \begin{tabular}[c]{@{}c@{}}\textbf{\textbf{\textbf{\textbf{Code}}}}\\\textbf{\textbf{\textbf{\textbf{(\%)}}}}\end{tabular} & \begin{tabular}[c]{@{}c@{}}\textbf{Doc}\\\textbf{\textbf{(\%)}}\end{tabular} \\ 
\midrule
    First Commit : & Development  & 31 & 98 & 2 \\
 & Repository Initializing & 43 & 77 & 23 \\
 & Re-engineering & 7 & 100 & 0 \\
 & Corrective Engineering & 2 & 100 & 0 \\
 & Management & 13 & 5 & 95 \\
 & Others & 4 & 100 & 0 \\ 
\midrule
    sum &  & 100 & \multicolumn{1}{l}{} & \multicolumn{1}{l}{} \\ 
\midrule
    Pull Request : & Development & 9 & 89 & 11 \\
 & Repository Initializing & 3 & 33 & 67 \\
 & Re-engineering & 17 & 76 & 24 \\
 & Corrective Engineering & 6 & 100 & 0 \\
 & Management & 60 & 45 & 55 \\
 & Others & 4 & 100 & 0 \\ 
\midrule
    sum &  & 100 &  &  \\
\bottomrule
\end{tabular}
\end{table}

\paragraph{\textbf{Frequency of Contribution's Kinds.}} Table~\ref{fig:rq2_graph} depicts the distribution for kinds of contributions made by \wang{\newcomers}. 
For the first commit contributions, as shown in the table,  31\% and 43\% of \newcomers~engage in development activities and repository initializing activities in the first commits. 
The result suggests that \newcomers~are more likely to engage in development activities (i.e., 31\% + 43\% = 74\%) when submitting first commits. 
Upon closer inspection, we find that 98\% and 77\% of development activities and repository initializing activities involve code related changes.
For the first PR contributions, our manual classification shows that 60\% of \newcomers~engage in management activities when submitting their PRs, indicating that \newcomers~are more likely to target maintenance activities.
Furthermore, we find that 45\% of management activities are related to formatting code, and 55\% are associated with cleaning up and updating documentation. 
More specifically, 4\% of their first commits and 4\% of first PRs contributions are classified as Others. Through our manual analysis, we find that these contributions are inaccessible (i.e., 404 errors), not be classified into any category based on our taxonomy, or not written in English. 

Our statistical tests confirm statistically significant differences between the proportion of development and maintenance activities for both types of contributions (first commit and PR), with a p-value $<$ 0.001.
For the type of first commit contributions, the test result validates that \newcomers~are more likely to engage in development activities.
However, for the type of first PR contributions, the test result confirms that \newcomers~are more likely to be involved in maintenance activities.
To conclude, our raised hypothesis, \textit{(H2) Contributions to GitHub repositories from \newcomers~are more likely to do development activities}, is established in first commit contributions, while it is not established in first PR contributions.

\begin{tcolorbox}
   \textbf{RQ2 Summary:}
    For the first commit contributions, we find that 74\% of contributions from \newcomers~are related to development activities.
    For the first PR contributions, our results show that 60\% of contributions are associated with management activities.
    Furthermore, statistical tests confirm that \textit{(H2) Contributions to GitHub repositories from \newcomers~are more likely to do development activities} is established in first commit contributions, but it is not established in first PR contributions.
\end{tcolorbox}

\label{sec:social_coding}
\subsection{Social Coding in Terms of Multiple Authorship (RQ3)} 
\begin{figure}[t]
    \centering
    \includegraphics[width=\linewidth]{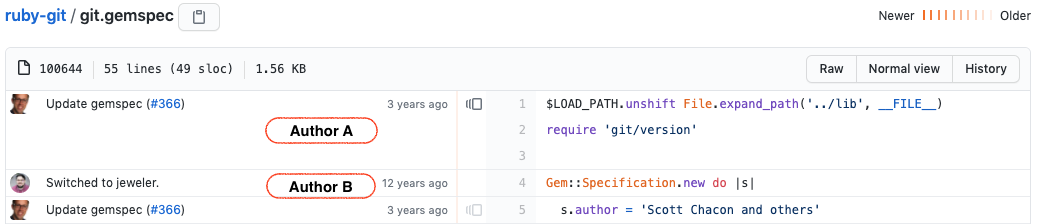}
     \caption{An example of how we define developers practice social coding, where more than one author contributes to the git.gemspec file.
     }.
    \label{fig:social_coding}
\end{figure}

\textit{Approach.} 
Social coding is a very loose term~\citep{dabbish2012social} used to describe the ability for developers to advertise (openly share and allow modification) their code on social platforms such as GitHub. In our paper, as shown in Figure~\ref{fig:social_coding}, we select one social coding practice in terms of multiple authorship to analyze where a contributor modifies either someone else's codes or others may modify this contributor's codes in the future. 
In the example, there are two authors (i.e., author A for lines 1--3 and author B for line 4) that contribute to a single file (i.e., git.gemspec) in a repository (i.e., ruby-git).
To do so, we use the \textit{D2} dataset from RQ2, which contains first commit and first PR contributions.
We identify social coding using Algorithm~1 and the \texttt{git-blame}\footnote{\url{https://www.atlassian.com/git/tutorials/inspecting-a-repository/git-blame}} command on each contained file in the commit to check whether the files receive changes from more than one author (lines 3--4 in Algorithm~1). 
Considering that one PR may include multiple commits, we analyze all commits inside each PR with Algorithm~1.
Specifically, we found that 21 out of 97 PRs (22\%) have multiple commits.

\begin{algorithm}[t]
  \caption{Identify social coding in terms of whether a contribution is modified by a single author or multiple authors.}
\footnotesize
    \SetKwInOut{Input}{Input}
    \SetKwInOut{Output}{Output}
    \SetKwInOut{Method}{Method}
    \Input{First \noindent $Commit$/$PR$ performed by an author $au$}
    \Output{Contribution type of First $Commit$/$PR$: single or multiple authors}
    $F \leftarrow$ A set of files modified by First $Commit$\,/$PR$\;
    $Type(F) = $single author\;
    \For{$f \in F$}{
    $D \leftarrow extract\_authors(\texttt{git-blame}(f))$\;
    \If{$au \in D$ \& $|D|$ $>$ 1}{
      $Type(f)$ = multiple author\;
    }
    }
    \Return $Type$;
  \label{Al1}
\end{algorithm}

To validate our hypothesis \textit{(H3) \newcomers~are more likely to contribute to a file with multiple authorship}. Similar to RQ1, we use the one proportion Z-test~\citep{z_test}.

\textbf{Social coding (Multiple Authorship).} Table~\ref{tab:rq1_stat} presents the frequency of social and non-social contributions in terms of authorship done by \newcomers. As shown in the table, the majority of \newcomers~do not practice social coding after joining GitHub.
For instance,  we find that 73\% of the first commits and 59\% of the first PRs are contributed by a single author.
Such results suggest that \newcomers~are less likely to practice social coding in terms of sharing multiple authorship, when placing their first GitHub contributions.

Our statistical test validates that for the first commits, there is a statistically significant difference between the proportion of social and non-social contributions, with a p-value $<$ 0.001, where \newcomers~are likely to practice non-social coding.
For the first PRs, there are no statistically significant difference, with a p-value $>$ 0.05.
To conclude, our proposed hypothesis \textit{(H3) \newcomers~are more likely to contribute to a file with multiple authorship}, is not established in both first commits and PRs.

\begin{table}[t]
	\fontsize{9}{11}\selectfont
	\tabcolsep=0.5cm
\centering
\caption{Frequency of social and non-social contributions from \newcomers~in terms of single/multiple authorship. After joining GitHub, 73\% and 59\% of \newcomers~have non-social based contributions in their first commits and PRs.}
\label{tab:rq1_stat}
\begin{tabular}{lrl}
\toprule
\textbf{Social coding practice (First Commit)} & \textbf{Percent (\%)} \\
\midrule
multiple &  27 & \mybar{0.27}   \\
single &  73 &\mybar{0.73}   \\
\midrule
\textbf{Social coding practice (Pull Request)} & \textbf{Percent (\%)} \\
\midrule
multiple &  41 & \mybar{0.41}   \\
single &  59 &\mybar{0.59}   \\
\bottomrule
\end{tabular}
\vspace{-.3cm}
\end{table}

\begin{tcolorbox}
    \textbf{RQ3 Summary:} Our results show that after joining GitHub, a majority of \newcomers~(i.e., 73\% of first commits and 59\% of PRs) do not share code with other authors. 
    Furthermore, statistical tests validate that \textit{(H3) \newcomers~are more likely to contribute to a file with multiple authorship}, is not established in both first commit and PR contributions. 
\end{tcolorbox}

\label{sec:onboarding}
\subsection{Onboarding of \newcomers~(RQ4)}

\textit{Approach.} To answer RQ4, we perform both quantitative and qualitative analyses. 
Different from the registered report, we find that making contributions to an OSS project is not trivial, and involves a process that follows two steps:
\begin{itemize}
    \item \textit{Fork an OSS repository}. The first step for any \newcomer~is to fork an OSS repository. 
Hence, we extracted 936 fork repositories out of a total of 2,392 repositories from the \textit{D1} dataset.
Then, to identify whether this repository is an engineered software project, we matched each fork repository against a curated dataset by \cite{meiESE2016}.
\item \textit{Identify contributions.} During step one, we found that many participants who only fork the repository, without contributing back to either the fork or upstream repository.
Hence, we performed an in-depth analysis through two particular ways of onboarding i.e., either the fork or upstream repositories.
\end{itemize}
    
For the qualitative analysis, we conducted a follow-up survey\footnote{Survey details are available at \url{https://forms.gle/JQiVamovUXdJiy8z5}} to acquire the perception of our participants. We sent our online survey invitation to \newcomers~through emails and ended up receiving 27 responses.
The survey is split into two questions, confirming whether participants had contributed to an OSS repository.
The first question is related to whether the participant had onboarded an OSS project (i.e., Since joining GitHub, did you successfully make a contribution to any Open Source Software project?).
In the second question, we explore the barriers faced by OSS newcomers \citep{Steinmacher2014}.
Hence, we asked participants to rate each barrier (i.e., Social Interaction, Newcomer Previous Knowledge, Finding a Way to Start, Technical Hurdles, and Documentation) on a five-point Likert scale.

\begin{table}[t]
\caption{Frequency of \newcomers~that started the onboarding process for OSS repositories.}
\tabcolsep=0.1cm
\centering
\label{tab:rq4_quan}
\begin{tabular}{llcc} 
\toprule
\begin{tabular}[c]{@{}l@{}}\textbf{Match to the}\\\textbf{Munaiah(2016) dataset}\end{tabular} & \multicolumn{1}{c}{\textbf{Onboarding Steps}} & \begin{tabular}[c]{@{}c@{}}\textbf{Count}\\\textbf{(\#)}\end{tabular} & \begin{tabular}[c]{@{}c@{}}\textbf{Percent}\\\textbf{(\%)}\end{tabular} \\ 
\midrule
Started Onboarding\\ Process : &  & 81 & 49 \\
\multicolumn{1}{l}{} & Fork an OSS repository (51\%) &  &  \\
\multicolumn{1}{l}{} & Contribute to fork OSS repository (22\%) &  &  \\
\midrule
Eventually Onboarded: & Contribute to original OSS repository (30\%) &  &  \\ 
\midrule
    Not Onboard:  &  & 85 & 51 \\ 
\midrule
    Sum &  & 166 & 100 \\
\bottomrule\\
\end{tabular}
\end{table}

\begin{figure}[ht]
    \centering
    \bigskip
    \begin{subfigure}{\columnwidth}
        \centering
        \renewcommand\tabularxcolumn[1]{m{#1}}
        \renewcommand\arraystretch{1.3}
        \setlength\tabcolsep{2pt}
    \label{qual_onboard_1}    
    \begin{tabular}{lcc}
    \toprule
    \begin{tabular}[c]{@{}l@{}}\textbf{Response to ``Since joining GitHub, did you successfully}\\\textbf{make a contribution to any OSS project?''}\end{tabular} & 
    \begin{tabular}[c]{@{}c@{}}\textbf{Count}\\\textbf{(\#)}\end{tabular}  & 
    \begin{tabular}[c]{@{}c@{}}\textbf{Percent}\\\textbf{(\%)}\end{tabular} \\
    \midrule
    Has made a contribution to an OSS project.   & 19 &  70\%    \\
    Has never made a contribution to an OSS project.   &  8  &  30\%  \\
    \midrule
        sum & 27 & 100 \\
    \bottomrule
    \end{tabular}
        \caption{Answers to Q1 in the follow-up survey.}
    \end{subfigure}
        \begin{subfigure}{\columnwidth}
        \label{qual_onboard_2}  
        \includegraphics[width=\linewidth]{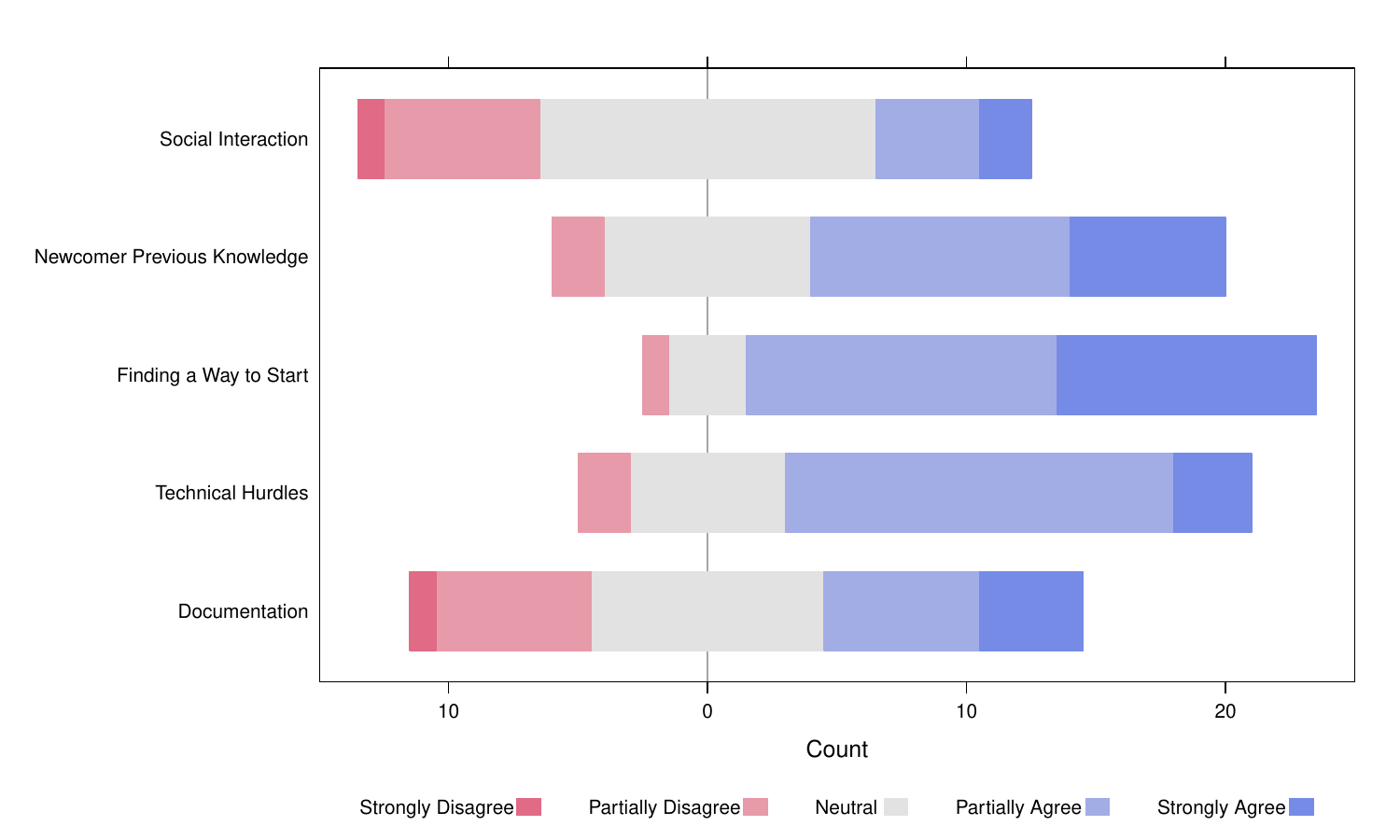}
    \caption{Barriers faced by \newcomers. Most \newcomers~(i.e., 22 out of 27 responses) strongly agree that \textit{finding a way to start} is a barrier.}
    \end{subfigure}
\caption{Qualitative analysis using a follow-up survey to acquire the perception of \newcomers.}
    \label{qual_all}
\end{figure}

{\textbf{Onboarding Process in GitHub.}}
Table~\ref{tab:rq4_quan} presents the distribution of how \newcomers~onboard OSS projects in terms of the quantitative analysis.
We show that 49\% of \newcomers~onboard OSS projects, while 51\% do not. Furthermore, 51\% of \newcomers~only fork the OSS repositories not making any contributions (Fork an OSS repository), and 22\% have contributed in the form of making commits to their own fork OSS repositories (Contributed to fork OSS repository). Meanwhile, 30\% of \newcomers~eventually onboard by submitting PRs directly to the original OSS repositories (Contributed to original OSS repository).
On the other hand, for the qualitative analysis, the survey results show that 19 out of 27 \newcomers~(70\%) claim that they have made contributions to OSS repositories. 
Figure 3 (a) shows the distribution of \newcomers~onboarding OSS projects by means of qualitative analysis. 


\textit{\textbf{Barriers faced by \newcomers.}}
Figure 3 (b) shows the results of our Likert-scale question related to barriers.
The figure shows that \textit{finding a way to start} is the most crucial barrier, with 22 responses being positive (i.e., 12 agree and 10 strongly agree responses). 
The second most crucial barrier is \textit{technical hurdles}, receiving 18 positive responses (i.e., 15 agree and 3 strongly agree responses).
\textit{Newcomer previous knowledge} is considered the third most crucial barrier with 16 responses (i.e., 10 agree and 6 strongly agree responses).  
On the other hand, the respondents are more likely to disagree with the statement that \textit{social interaction} and \textit{documentation} can be barriers for them to onboard OSS projects (i.e., 7 negative responses for each barrier).

\begin{tcolorbox}
    \textbf{RQ4 Summary:} 
    Our quantitative analysis shows that 30\% of \newcomers~eventually onboarded OSS projects.
    Our follow-up user survey also shows that 19 out of the 27 participants (70\%) claim that they have made contributions to OSS repositories.
    We find that \textit{finding a way to start} is the most agreed barrier for \newcomers.
\end{tcolorbox}


\section{Discussions}
\label{sec:discussion}
In this section, we discuss deviations from the registered report, lessons learned and then revisit our expected implications listed in the registered report against the actual results.

\subsection{Deviations}

The execution of this registered report (RR) prompted unavoidable changes to our protocols. We list up the following four deviations below: \textit{(i) Term Newcomer OSS-Candidate.} To generalize the definition of the term, Newcomer Candidate has been changed to Newcomer OSS-Candidates as ``a developer that does not have any prior experience contributing to an OSS project, is a new user to a social coding platform, with the intention to onboard an OSS project'', \textit{(ii)Terminology Clarification.} In the registered report, our preliminary study is now a separate section in the full study. In terms of clarity, in the executed study, we specify the social coding practice as the number of authors on a shared file, and realize that onboard is an ongoing process, \textit{(iii) Research Design.} The statistical test has been changed to one proportion Z-test (Paternoster et al, 1998). After revising the categories, we realized that the statistical test in the RR was not appropriate. We modified the statistical tests based on the binary result categories of RQ1, RQ2, and RQ3. The one proportion Z-test compares an observed proportion to a theoretical one when the categories are binary, and last \textit{(iv) Hypothesis.} We adjusted the hypotheses H2 and H3. For H2, we changed it to \textit{(H2) Contributions to GitHub repositories from Newcomer OSS-Candidates are more likely to do development activities}, to be aligned with our motivation. For H3, we narrowed down the aspect of social coding and adjusted it to \textit{(H3) Newcomer OSS-Candidates are more likely to contribute to a file with multiple authorship.}

\subsection{Lessons learned}
This paper discusses two lessons learned that would be useful for future replication or improvements of the study.
In the first lesson, we acknowledge that extracting the first contribution is not as trivial as we first envisioned.
This is because the actual first commit might be just an ad-hoc test for the user, and not an actual meaningful contribution to a repository.
In this research, we manually filtered out such contributions, but future work should consider a more systematic approach.

The second lesson to acknowledge is the process of onboarding may take a long time as it may be tied with the process of making a contribution to GitHub.
As shown in the results for RQ4, different Newcomer OSS-Candidates are at different stages of the onboarding process and may take time before they decide to submit the PR.
Thus, we need to take into consideration a long enough time-window to evaluate whether or not a Newcomer OSS-Candidate will end up onboarding an OSS project.
\subsection{Implications (Expectations vs. Actual Results)}
Based on our results, we revisit our expected implications against the actual results of the study.\\

\textbf{Suggestions for Newcomers.} 
In our registered report, we speculated that our research would help \newcomers~understand the kinds of contributions they target before onboarding a real OSS project. 
Actually, we found in Table~\ref{fig:rq2_graph} that \newcomers~are not only engaged in adding new content, but 60\% of them are also interested in management activities related to formatting code, cleaning up, and updating documentation through the submission of PRs. 
One example of this can be seen in the AEOL's repository\footnote{\url{https://github.com/AE0L/round-robin/pull/1}}, where a PR is submitted to add a new function to the project.
Furthermore, RQ2 also reveals that after joining GitHub, 43\% of \newcomers~prefer to add new content in order to initialize or start a repository in their first commit.
We found a common pattern is an initial commit that is uploading a website to the GitHub repository.\footnote{\url{https://github.com/maanizfar/vanilla-js-web-projects/commit/e208d861b80762be8aa545567a300a7fad6aacf7}} Finally, based on our RQ3 quantitative analysis, the majority of \newcomers~have non-social based contributions in their contributions. As shown in Table~\ref{tab:rq1_stat} from RQ3 that
after joining GitHub, \newcomers~contributes in terms of single authorship are 73\% of their first commits and 59\% of their PRs, respectively.
On the basis of evidence, we conclude that it is unlikely that \newcomers~will be onboard to OSS projects immediately after joining GitHub.

We also speculated that we would reveal barriers on why some \newcomers~never end up contributing to an OSS projects. 
According to our survey responses in RQ4, finding a way to start is one of the most challenging barriers, with 22 responses being positive (i.e., 12 agree and 10 strongly agree responses). 
Hence, inspired by these examples and combining all results, we recommend that \newcomers~should not be afraid to individually contribute to their own code, contribute to upstream software repositories, or fork OSS projects before attempting to onboard. 
Last, regarding the most challenging barrier (i.e., finding a way to start), to this end, \newcomers~should leverage suggestions provided by \cite{Subramanian2020IEEE}, including minor feature additions (a change of around 36 lines of code), minor documentation changes, selecting bug fixes, and changing catering to revised dependencies as first-timer friendly, which may relieve this problem. 
In addition, there are online resources\footnote{\url{https://hacktoberfest.digitalocean.com/}} that help \newcomers~choose easy issues or opportunities to find ways to start contributing.

\textbf{Suggestions for OSS Projects.} 
The registered report speculated that the findings would reveal insights into what contributions may attract a~\newcomer. 
Through our qualitative analysis of RQ2, Table~\ref{fig:rq2_graph} shows that in the first commits, 43\% of \newcomers~are typically engaged in adding new content to initialize the repositories, and 60\% are involved in management activities in their PRs. 
Hence, we suggest that \newcomers~may not have required skills to make immediate contributions. 
Instead, they may start with software based upstream experimental repositories. Hence, for OSS project, it might start with tasks to update the documentation, formatting or cleaning up code. One example of this can be seen in Bviveksingh's upstream repository\footnote{\url{https://github.com/Bviveksingh/angular-starter/pull/1}}, where a PR is submitted to update a software version.

We also speculated that OSS projects may benefit from our study, by identifying and offering the right contributions for the right Newcomer OSS-Candidates.
Based on the results, we could not be able to provide concrete examples of contributions that match a specific Newcomer OSS-Candidate as the majority is a mixture of management and development activities.
A potential future venue for research could be to explore the kinds of OSS projects that these Newcomer OSS-Candidates end up onboarding. 
This would provide insights into matching the contributions to the onboarded OSS projects.

\textbf{Suggestions for Researchers.} 
The registered report speculated that non-software repositories that are personal have always been regarded as a challenge and are often filtered out from the dataset.
We find that the majority of targeted repositories are software based repositories. 
Results include experimental (24\%), documentation (21\%), and web-based-application-libraries-and-frameworks (17\%).
For researchers, this insight helps to understand the role of software based experimental, documentation, and web-based-application-libraries-and-frameworks repositories in platforms like GitHub, that should cater for developers.
A potential avenue for research is to perform a finer-grain of analysis to understand the nature of these repositories.

\section{Threats to Validity}
\label{sec:Lim_threat}
In this section, we now discuss threats to the validity of our study.

\textbf{External Validity.} Two external threats are identified. We perform an empirical study on \newcomers~that use GitHub the platform, and our observations may not be generalized to other platforms. 
Hence, we use GitHub as a case study. 
Another external threat is whether or not the 171 participants are representative of all \newcomers~of the GitHub platform. Hence, we rely on the first contribution community.
To represents the global population, future work should be conducted with other communities.

{\textbf{Construct Validity.}} We summarize three threats regarding construct validity.
First, our qualitative analysis of manually classifying repositories and contribution kinds (RQ1, RQ2) are prone to error.
To mitigate this threat, we took a systematic approach to first test our comprehension with 30 samples using Kappa agreement scores with three separate individuals.
The second threat is to identified first contributions in RQ2 may not be actual contributions. To mitigate this, we perform a manual inspection to ignore any test, not meaningful contributions (i.e., commits or PRs) from any experimental repositories.
The third potential threat exists in the quantitative analysis of matching engineered software projects using the curated database provided by \cite{meiESE2016}. 
We did contact the authors for assistance to help run the latest scripts, but were unsuccessful. 
Although the curated database might be outdated, we are confident that with the dataset, we were able to match 936 repositories.

{\textbf{Internal Validity.}} We identify three internal threats. The first threat is the first contributions by~\newcomers~may not be meaningful; they just want to get into the GitHub way of doing things. To mitigate this, we applied our first filter.
The second internal threat to validity is related to results obtained from the quantitative analysis of RQ3 adapted to data visualization. As per the result, 27\% and 41\% of social coding is done by~\newcomers~in their first commits and PRs. 
The final threat is regarding errors in our tracking of repositories, due to repositories being deleted or a user changing user ids, as studied by \cite{wiese2016mailing}. We acknowledge this threat, however, during our manual inspection, we are confident that this was only for a few cases.

\section{Related Work}
\label{sec:Related_Work}
A steady of influx of new developers to an OSS project is crucial for its sustainability. 
In this section, we compare and contrast our work to the prior studies in three parts: first, we introduce the studies that are related to motivation for newcomers and OSS projects; second, we consider the studies regarding onboarding OSS projects; third, we discuss the studies with respect to the barriers that newcomers face.

\textbf{Studies on Onboarding Motivators.}
There is a complete body of work that explored OSS developer's motivation and project's attractiveness \citep{Meirelles2010SBES,Santos2013strareg,shah&sonali2006MS,ye2003ICSE}. 
Studies have also investigated the progression from newcomer to a core project member \citep{Ducheneaut2005SocializationIA,Fang2009JMIS,Krogh2003RP,marlow2013CSCW,Nakakoji2003IWPSE}. 
On the other hand, \cite{choi2010socialization} identified the seven most frequently used socialization tactics which have impact on newcomers' commitment to online groups.
Other parts of the literature focus on the forces of motivation and attractiveness that drive newcomers towards projects. 
For example, \cite{Lakhanikarim2003whyhakers} have found that external benefits (e.g., better jobs, career advancement) motivate primarily new contributors, along enjoyment-based intrinsic, code-based challenges, and improving programming skills. 
Compared to these, our study investigates how \newcomers~contribute to both software (e.g., experimental, documentation, and web-based-application-libraries-and-frameworks ) and non-software (e.g., academic, Web, and storage) repositories.
Different to prior work, our goal is to study potential \newcomers~that have the intention to onboard an OSS project.

\textbf{Studies on the Onboarding Process}
There have been several studies that investigated the onboarding process.
\cite{Fagerholm2013ICGSE} presented preliminary observations and results of in-progress research that studied the process of onboarding into virtual OSS teams. Commercial software development settings are also affected by newcomers onboarding towards OSS projects, as described by \citet{Dagenais2010ACM,BegelICER2008}. \cite{Ducheneaut2005SocializationIA} approached onboarding from a sociological point of view by considering the perspective of individual developers. 
Previously, mentorship activity is recognized as an important factor for effective onboarding of newcomers towards OSS projects~\citep{Fagerholm2013ICGSE,fagerholm2014role,musicant2011mentoring}.
\cite{Swap2001MIS} described mentoring in their study as a basic knowledge transfer mechanism in the enterprise. 
A joining script is proposed in another study by \cite{Krogh2003RP} for developers who want to take participate in OSS project. 
\cite{Nakakoji2003IWPSE} also studied the OSS project and proposed eight possible joining roles comprise of concentric layers called ``the onion patch''. 
\cite{Zhou2015TSE} found that the willingness of individual and project’s climate were associated with odds that an individual would become a long-term contributor. 
Different from previous research, our study looks at the activities of potential newcomers before they onboard.

\textbf{Studies on the barriers to Onboarding.}
Newcomers are important to the survival, long-term success, and continuity of OSS projects~\citep{Kula2019book}. 
However, newcomers face many difficulties when making their first contributions to a project.
According to~\cite{ye2003ICSE}, learning is one of the motivational forces that motivates people to participate in OSS communities. Conversely, newcomers to a project send contributions which are not incorporated into the source code and give up trying~\citep{Igor2015CSCW}. 
As discussed by~\cite{zhou2010}, the transfer of entire projects to offshore locations, aging and renewal of core developers in legacy products, recruiting in fast growing Internet companies, and the participation in open source projects, presents similar challenges of rapidly increasing newcomer competence in software projects.
Several research activities are performed to reduce the barriers for newcomers previously. \cite{Steinmacher2014} proposed a developer joining model that represents the stages that are common and the forces that are influential to newcomers being drawn or pushed away from a project. \cite{Steinmacher2016ICSE} created a portal called FLOSScoach based on a conceptual model of barriers to support newcomers. The evaluation shows that FLOSScoach played an important role in guiding newcomers and in lowering barriers related to the orientation and contribution process. 
In terms of barriers, our research complements the work of \cite{Steinmacher2014}, which highlights the most crucial barrier among others, i.e., finding a way to start, due to which newcomers face difficulty in contributing to OSS projects.
Furthermore, our work takes a first look at potential \newcomers~before they onboard.
Hence, insights show that learning the social platform contribution process (i.e., PR process) may co-inside with onboarding.

\section{Conclusion}
\label{sec:conclusion}
In this work, we studied the activities of a particular category of potential contributors (i.e., Newcomer OSS-Candidates) towards OSS projects on GitHub.
To do that, we (i) analyze what kinds of repositories they target, (ii) investigate what kinds of contributions come from them, (iii) analyze to what extent they practice social coding with their contributions, and (iv) explore what proportion of them eventually onboard an OSS project.

We observe that (i) 66\% of Newcomer OSS-Candidates target software based repositories;
(ii) the majority of their contributions are related to development activities and maintenance activities, respectively, for commits and PRs; (iii) Newcomer OSS-Candidates are less likely to practice social coding in their contributions in terms of multiple authorship; and (iv) 70\% of them eventually onboarded OSS projects in a follow-up survey and cited that \textit{finding a way to start} is the most crucial barrier.
As GitHub continues to grow, so does the possibility to attract potential contributors to OSS projects.
Our work presents the first step towards understanding these potential contributors and reveals insights to provide a guidance for them to onboard an OSS project.
\\

\section*{Acknowledgement}
This work is supported by Japanese Society for the Promotion of Science (JSPS) KAKENHI Grant Numbers 18H04094 and 20K19774 and 20H05706.

\bibliographystyle{spbasic}      
\bibliography{bibliography.bib}   

\end{document}